\documentclass[sn-aps]{sn-jnl}

\usepackage{graphicx}%
\usepackage{multirow}%
\usepackage{amsmath,amssymb,amsfonts}%
\usepackage{amsthm}%
\usepackage{mathrsfs}%
\usepackage[title]{appendix}%
\usepackage{xcolor}%
\usepackage{textcomp}%
\usepackage{manyfoot}%
\usepackage{booktabs}%
\usepackage{algorithm}%
\usepackage{algorithmicx}%
\usepackage{algpseudocode}%
\usepackage{listings}%

\theoremstyle{thmstyleone}%
\theoremstyle{thmstyletwo}%

\theoremstyle{thmstylethree}%

\begin{document}
\title[Fluctuations in email size modeled using a gamma-like distribution]{Fluctuations in email size modeled using a gamma-like distribution}


\author*[1]{\fnm{Yoshitsugu} \sur{Matsubara}}\email{matubara@cc.saga-u.ac.jp}



\affil*[1]{\orgdiv{Computer and Network Center}, \orgname{Saga University}, \orgaddress{\street{1 Honjo-machi}, \city{Saga-shi}, \postcode{840-8502}, \state{Saga}, \country{Japan}}}




\abstract{
A previously established frequency distribution model, which integrates a lognormal distribution with a logarithmic equation, effectively characterizes fluctuations in email size during sending requests.
In addition, an email size generation model has been developed based on this log-normal-like framework.
While the fitting of these models has been deemed satisfactory, they can be further enhanced in the range of small email sizes.
This study advances these models by incorporating a gamma distribution alongside a logarithmic equation.
The resulting gamma-like model demonstrates a significantly improved fit compared with the log-normal-like model.
These results contribute to the knowledge on the statistical properties of sending mail.
}

\keywords{Email size, Power-law fluctuations, Gamma-like distribution Log-normal-like distribution}

\maketitle
\section{Introduction} \label{sec1}
The complex structures and dynamic behaviors of Internet systems have garnered significant attention in recent years.
Researchers have extensively investigated phenomena such as scale-free structures~\cite{Faloutsos:1999} and power-law distributions in packet flows~\cite{Paxson:1995,Csabai:1994,Takayasu:1996,Tadaki:2007,Eckmann:2004,Barabasi:2005,K.-I.:2008,Malmgren:2008,Anteneodo:2010,Karsai:2012}, particularly during inter-event times.
Investigations have also focused on power-law correlations in data flow and email sending requests~\cite{Matsubara:2013} during these intervals.

A previous study identified power-law fluctuations in email size during send requests~\cite{Matsubara:2017} and proposed a novel model based on a log-normal-like distribution.
This model effectively captured the power-law characteristics associated with large emails and has been referenced in numerous subsequent studies~\cite{ASI:ASI21331}.
An email size generation model elucidated the mechanisms underlying log-normal-like distributions~\cite{Matsubara:2025-1}, with linguistic principles applied to explain the foundational aspects of this model further.

However, the model did not adequately fit the observed data in the small email size range.
This limitation arises because the size generation model assumes that the length of a single word in an email’s content follows a normal distribution, which permits a negative length for a single word.
Given that the minimum size of a single word must be at least one byte, we propose a gamma distribution characterized by a logarithmic equation to enhance the log-normal-like model.
The simulation results from the gamma-like frequency distribution model demonstrated stronger agreement with the observed frequency distributions.


\section{Email size fluctuations}\label{sec2}

The frequency distribution of email sizes across various periods and user groups from May 1, 2015 to July 31, 2015 was examined~\cite{Matsubara:2017}.
The analysis revealed two significant inflection points in the size-frequency distribution, occurring at approximately $15 \,\mathrm{kB}$ and $40 \,\mathrm{kB}$.
The analysis focused on emails with attachments, as plain-text emails typically remained below several tens of kilobytes in size.
To enhance the understanding of the data, the size-frequency distribution was categorized into two subdistributions based on content type: ``no attachment'' and ``attachment.''
The ``no attachment'' category encompasses both plain-text and HTML emails.
The content types were classified in accordance with the Multipurpose Internet Mail Extension (MIME) protocol~\cite{rfc2045,rfc2046,rfc2047,rfc2049,rfc4289,rfc6838}.

Given that email bodies comprise written sentences, linguistic principles were incorporated into the analyses.
Most users in the organization from which the data were sourced were Japanese, resulting in emails that were predominantly composed in Japanese and English.
Previous studies that analyzed sentence length have identified various distribution types, including log-normal~\cite{Arai:2001,Furuhashi:2012}, gamma~\cite{Sasaki:1976}, and hyper-Pascal~\cite{Ishida:2007} distributions, among others~\cite{borbely2019sentencelength,Liu_Yang_Liu_2024}.

Based on this analysis, the frequency distribution model is characterized by a probability density function $p_{\mathrm{LNL}}(s)$ for each subdistribution of email size frequencies, where $s$ is measured in units of 100 bytes, as outlined in~\cite{Matsubara:2017}:
\begin{eqnarray}
p_{\mathrm{LNL}}(s) & = & \frac{1}{a}\frac{1}{s\ln{s}} \exp{ \left\{ \frac{- (\ln{\ln{s}} - \mu)^{2}}{2 \sigma^{2}} \right\} } \label{model_eq_log-normal-like},
\end{eqnarray}
where $a$ represents the normalized constant ($a > 0$), $-\infty < \mu < \infty$, and $\sigma > 0$\footnote{The function is continuous over the range  ($1,\infty$) for $s$ for $s$. Eq.~\ref{model_eq_log-normal-like} integrates the log-normal distribution $g(x) = \frac{1}{a}\frac{1}{x}\exp(-\frac{(\ln{x} - \mu)^{2}}{2\sigma^{2}})$ with $x = \ln{s}$.}.
The expected value of $p_{\mathrm{LNL}}(s)$ remains unknown.
The logarithms in eq..~\ref{model_eq_log-normal-like} for $s \gg 1$ can be approximated as follows:
\begin{eqnarray}
\ln{p_{\mathrm{LNL}}(s)} & \sim & - \ln{s} - \frac{(\ln{\ln{s}})^{2}}{2 \sigma^{2}} \nonumber \\
& \sim & - \ln{s}, \nonumber
\end{eqnarray}
where $g(x)$ does not exhibit power-law behavior for  $x \gg 1$.
Consequently,  $x = \ln{s}$ influences the power-law properties of $p_{\mathrm{LNL}}(s)$ when $s \gg 1$.
The power law indicates that setting an upper limit on the sendable email size is beneficial for email services.
One of the qualitative reasons for the upper limit is to reduce the load on email services.
If the frequency of the email size decreases exponentially, it can be inferred that huge size emails overload the mail service are seldom sent.
If this is the case, there may be no practical problem for the email service recognizing an unlimited size of sendable emails.
However, this power law suggests that huge email sizes overload the mail service should be considered.
Therefore, this power-law property provides useful knowledge for the operation of email services.

The proposed size generation model  $s_{t}$ at time $t$ is expressed as follows~\cite{Matsubara:2025-1}:
\begin{equation}
s_{t} = b s_{t-1}^{c} \mathrm{e}^{\mathrm{e}^{\epsilon_{t}}}, \label{model_eq_email-size-generation}
\end{equation}
where $b$ and $c$ represent coefficients and $\epsilon_{t}$ represents the length of a single word following a normal distribution  $\mathcal{N}(\mu,\sigma^{2})$.
$\exp\{\epsilon_{t}\}$ represents the length of a single sentence and $\exp\{\exp\{\epsilon_{t}\}\}$ represents the length of a compound sentence.


\section{Gamma-like frequency distribution and email size generation model} \label{propose}

In eq.~\ref{model_eq_email-size-generation}, the term  $\epsilon_{t}$ represents the length of a single word within an email text, and its distribution is modeled as a normal distribution.
Notably, the length of a single word must be at least one byte.
However, the normal distribution allows for the possibility of negative values for word length.
To address this limitation, a gamma-like frequency distribution model  $p_{\mathrm{GL}}(s)$ is proposed as an enhancement to eq.~\ref{model_eq_log-normal-like}:
\begin{eqnarray}
p_{\mathrm{GL}}(s) & = & \frac{1}{a}\frac{1}{s\ln{s}} \frac{1}{\theta^{k}} (\ln{\ln{s}})^{k-1} \exp{ \left\{ -\frac{\ln{\ln{s}}}{\theta} \right\}}, \label{model_eq_gamma-like}
\end{eqnarray}
where  $s > \mathrm{e}$.
The expected value of  $p_{\mathrm{GL}}(s)$ also remains unknown.

If  $s \gg 1$, the logarithm of both sides of eq.~\ref{model_eq_gamma-like} can be approximated as follows:
\begin{eqnarray}
\ln{p_{\mathrm{GL}}(s)} & \sim & - \ln{s} - (1+\frac{1}{\theta}) \ln{\ln{s}} + (k-1) \ln{\ln{\ln{s}}} \nonumber \\
& \sim & - \ln{s}. \nonumber
\end{eqnarray}
This model also represents the power-law property in $s \gg 1$.

The term $\epsilon_{t}$ in the size generation model (eq.~\ref{model_eq_email-size-generation}) transitions from a normal distribution $\mathcal{N}(\mu,\sigma^{2})$ to a gamma distribution $(k,\theta)$.
This transition resolves the negative length for a single word.


\section{Discussion} \label{Discussion}

\begin{figure}[t]
\centering
\resizebox{1.00\textwidth}{!}{\includegraphics{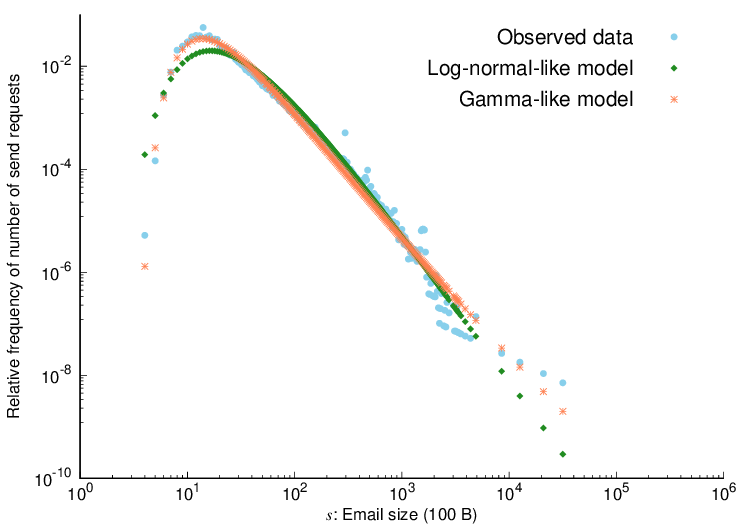}}
\caption{
Size-frequency distribution of the ``no attachment'' emails in the observed data, log-normal-like frequency model $p_{\mathrm{LNL}}(s)$, and gamma-like frequency model $s_{\mathrm{GL}}(s)$.
The frequency distribution is logarithmically binned~\cite{ASI:ASI21426}, with both axes on a logarithmic scale.
The blue points represent the observed relative frequency values, the green points denote those calculated using the log-normal-like frequency, and the orange points represent those calculated using the gamma-like frequency.
The horizontal axis represents the email size $s$ (in units of 100 bytes), ranging from $0 \, \mathrm{MB}$ to $10 \, \mathrm{MB}$, with the bin size $\Delta s$ defined as $(10^{0.01} - 1) s$s
The sizes of all bin intervals were consistent on the logarithmic scale.
The vertical axis denotes the relative frequency of emails.
The degrees of fitting were $D = 88.8502$ and $70.3002$ for the log-normal-like and gamma-like distributions, respectively.
}
\label{res_size_frequency_per100B_noattachment_log10_1}
\end{figure}

The fitting results for  $p_{\mathrm{LNL}}$ and $p_{\mathrm{GL}}$ are shown in fig.~\ref{res_size_frequency_per100B_noattachment_log10_1}.
The least-squares method, denoted by $D$, was utilized to assess the goodness-of-fit~\cite{Matsubara:2017}.
$D$ is defined as $D = \sum_{s} (\ln{y_{O}(s)} - \ln{y(s)})^2$2, where$y_{O}(s)$ represents the relative frequency of each bin size in the observed data and $y(s)$ represents the relative frequency derived from the proposed model.
$D$ approaches zero when $y_{O}(s) \simeq y(s)$.

The degree of fitting in  $p_{\mathrm{LNL}}$ was $D = 88.8502 \, (\mu = 0.0489, \sigma = 0.2461)$, whereas that in $p_{\mathrm{GL}}$ was $D = 70.3002 \, (k = 24.1803, \theta = 0.0489)$.
Notably, the gamma-like distribution $p_{\mathrm{GL}}$ achieved superior results compared with the log-normal-like distribution $p_{\mathrm{LNL}}$.

\begin{figure}[t]
\centering
\resizebox{1.00\textwidth}{!}{\includegraphics{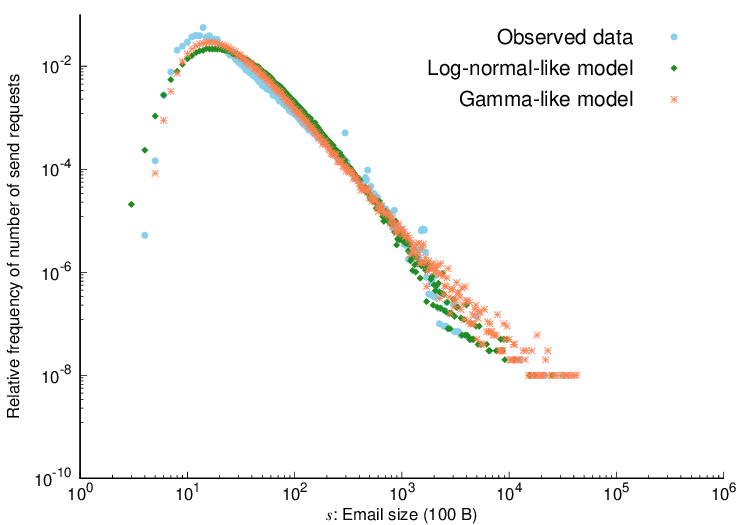}}
\caption{
Size-frequency distribution of the ``no attachment'' emails in the observed data, those of the size generation model with a log-normal-like frequency, and those of the size generation model with a gamma-like frequency.
The blue points represent the observed relative frequency values, the green points represent those calculated using the log-normal-like frequency, and the orange points represent those calculated using the gamma-like frequency.
The degrees of fitting were $D = 15.19232$ and $13.99222$ for the log-normal-like and gamma-like distributions, respectively.
}
\label{res_size_frequency_per100B_noattachment_simulation_log10_1}
\end{figure}

The email size frequency distribution generated using st was simulated.
The results are shown in fig.~\ref{res_size_frequency_per100B_noattachment_simulation_log10_1}.
This study generated 191,993 emails based on st, aligning with the number of ``no attachment'' emails analyzed in a previous study~\cite{Matsubara:2017}.
The parameter values of $p_{\mathrm{LNL}}$ were $\mu = 1.259$, $\sigma = 0.235$, whereas those of $p_{\mathrm{GL}}$ were $k = 24.1803$, $\theta = 0.0489$.
The degrees of fit were $D = 15.19232$ and $13.99222$ for $p_{\mathrm{LNL}}$ and $p_{\mathrm{GL}}$, respectively, indicating that the frequency distribution of the gamma-like model, $p_{\mathrm{GL}}$, is a better fit than that of the log-normal-like model, $p_{\mathrm{LNL}}$.

%
%
%
%
%


\section{Conclusion} \label{conclusion}

A previous study investigated the frequency distributions of email sizes in sent requests~\cite{Matsubara:2017} and proposed a log-normal-like model, denoted as $p_{\mathrm{LNL}}(s)$.
This model integrates the lognormal distribution $g(x)$ and  $x = \ln{s}$, where $s$ denotes the email size.
The proposed model effectively approximates power-law correlations when $s \gg 1$.
In addition, a model for email size generation was proposed to elucidate the mechanism underlying the size–frequency distribution $p_{\mathrm{LGL}}$~\cite{Matsubara:2025-1}.

An enhanced model, referred to as $p_{\mathrm{GL}}(s)$, was developed to improve upon $p_{\mathrm{LNL}}(s)$.
This new model combines the gamma distribution $f(x)$ and $x = \ln{\ln{s}}$.
Similarly,  $p_{\mathrm{GL}}(s)$ denotes power-law correlations when $s \gg 1$.
Notably, the frequency distribution $p_{\mathrm{GL}}(s)$ demonstrated a superior fit compared with that of  $p_{\mathrm{LNL}}(s)$.
Furthermore, the frequency distribution generated based on $p_{\mathrm{GL}}(s)$ outperformed that of $p_{\mathrm{LNL}}(s)$.

%
%

Each relationship of each term (a single word $\epsilon_{t}$, single sentence, and compound sentence) in the size generation model should be validated by actual email content data.
Notably, the contents of the emails were not analyzed owing to the challenges associated with third-party access to email contents, which are protected by privacy regulations.
Analyzing email contents would be beneficial for validating and refining this model; however, this aspect has been designated for future research.

\backmatter
\bmhead{Compliance with ethical standards}

The authors declare no competing interests relevant to the contents of this paper.





\bmhead{Data Availability Statement}

The data are not publicly available to avoid compromising information privacy.


\bmhead{Acknowledgments}

The author expresses gratitude to Makoto Otani from the Computer and Network Center, Saga University, for his assistance with data collection and analysis.

\bibliography{matsubara,rfc}

\begin{thebibliography}{10}
\providecommand{\url}[1]{{#1}}
\providecommand{\urlprefix}{URL }
\providecommand{\doi}[1]{\url{https://doi.org/#1}}
\bibcommenthead

\bibitem{Faloutsos:1999}
M.~Faloutsos, P.~Faloutsos, C.~Faloutsos, On power--law relationships of the
  internet topology.
\newblock SIGCOMM Comput. Commun. Rev. \textbf{29}(4), 251–262 (1999).
\newblock \doi{10.1145/316194.316229}

\bibitem{Paxson:1995}
V.~Paxson, S.~Floyd, Wide area traffic: the failure of {Poisson} modeling.
\newblock IEEE/ACM Trans. Networking \textbf{3}, 226--244 (1995).
\newblock \doi{10.1109/90.392383}

\bibitem{Csabai:1994}
I.~Csabai, 1/f noise in computer network traffic.
\newblock J. Phys. A: Math. Gen. \textbf{27}(12), L417 (1994).
\newblock \doi{10.1088/0305-4470/27/12/004}

\bibitem{Takayasu:1996}
M.~Takayasu, H.~Takayasu, T.~Sato, Critical behaviors and $1/f$ noise in
  information traffic.
\newblock Physica A \textbf{233}, 824--834 (1996).
\newblock \doi{10.1016/S0378-4371(96)00189-6}

\bibitem{Tadaki:2007}
S.~Tadaki, Power-law fluctuation in internet traffic.
\newblock J. Phys. Soc. Jpn. \textbf{76}(3), 044001--044001--5 (2007).
\newblock \doi{10.1143/jpsj.76.044001}

\bibitem{Eckmann:2004}
J.P. Eckmann, E.~Moses, D.~Sergi, Entropy of dialogues creates coherent
  structures in e-mail traffic.
\newblock Proc. Nattl. Acad. Sci. U.S.A. \textbf{101}(40), 14333--14337 (2004).
\newblock \doi{10.1073/pnas.0405728101}

\bibitem{Barabasi:2005}
A.L. Barab\'{a}si, The origin of bursts and heavy tails in human dynamics.
\newblock Nature \textbf{435}, 207--211 (2005).
\newblock \doi{10.1038/nature03459}

\bibitem{K.-I.:2008}
K.I. Goh, A.L. Barab\'{a}si, Burstiness and memory in complex systems.
\newblock EPL (Europhys. Lett.) \textbf{81}(4), 48002 (2008).
\newblock \doi{10.1209/0295-5075/81/48002}

\bibitem{Malmgren:2008}
R.D. Malmgren, D.B. Stouffera, A.E. Motter, L.A.N. Amaral, A {Poissonian}
  explanation for heavy tails in e-mail communication.
\newblock Proc. Natl. Acad. Sci. U.S.A. \textbf{105}(47), 18153--18158 (2008).
\newblock \doi{10.1073/pnas.0800332105}

\bibitem{Anteneodo:2010}
C.~Anteneodo, R.D. Malmgren, D.R. Chialvo, Poissonian bursts in e-mail
  correspondence.
\newblock Eur. Phys. J. B \textbf{75}, 389--394 (2010).
\newblock \doi{10.1140/epjb/e2010-00139-9}

\bibitem{Karsai:2012}
M.~Karsai, K.~Kaski, A.L. Barab\'{a}si, J.~Kert\'{e}sz, Universal features of
  correlated bursty behaviour.
\newblock Sci. Rep. \textbf{2}, 397 (2012).
\newblock \doi{10.1038/srep00397}

\bibitem{Matsubara:2013}
Y.~Matsubara, Y.~Hieida, S.~Tadaki, Fluctuation in e-mail sizes weakens
  power-law correlations in e-mail flow.
\newblock Eur. Phys. J. B \textbf{86}, 209 (2013).
\newblock \doi{10.1140/epjb/e2013-40209-x}

\bibitem{Matsubara:2017}
Y.~Matsubara, Y.~Musashi, Fluctuations in email size.
\newblock Eur. Phys. J. Plus \textbf{132}, 507 (2017).
\newblock \doi{10.1140/epjp/i2017-11767-2}

\bibitem{ASI:ASI21331}
S.~Milojevi\v{c}, Modes of collaboration in modern science: Beyond power laws
  and preferential attachment.
\newblock Journal of the American Society for Information Science and
  Technology \textbf{61}(7), 1410--1423 (2010).
\newblock \doi{10.1002/asi.21331}

\bibitem{Matsubara:2025-1}
Y.~Matsubara.
\newblock Fluctuations in the email size modeled by a log-normal-like
  distribution (2025).
\newblock \doi{10.48550/arXiv.2501.04042}

\bibitem{rfc2045}
N.~Freed, D.N.S. Borenstein.
\newblock {Multipurpose Internet Mail Extensions (MIME) Part One: Format of
  Internet Message Bodies}.
\newblock RFC 2045 (1996).
\newblock \doi{10.17487/RFC2045}

\bibitem{rfc2046}
N.~Freed, D.N.S. Borenstein.
\newblock {Multipurpose Internet Mail Extensions (MIME) Part Two: Media Types}.
\newblock RFC 2046 (1996).
\newblock \doi{10.17487/RFC2046}

\bibitem{rfc2047}
K.~Moore.
\newblock {MIME (Multipurpose Internet Mail Extensions) Part Three: Message
  Header Extensions for Non-ASCII Text}.
\newblock RFC 2047 (1996).
\newblock \doi{10.17487/RFC2047}

\bibitem{rfc2049}
N.~Freed, D.N.S. Borenstein.
\newblock {Multipurpose Internet Mail Extensions (MIME) Part Five: Conformance
  Criteria and Examples}.
\newblock RFC 2049 (1996).
\newblock \doi{10.17487/RFC2049}

\bibitem{rfc4289}
D.J.C. Klensin, N.~Freed.
\newblock {Multipurpose Internet Mail Extensions (MIME) Part Four: Registration
  Procedures}.
\newblock RFC 4289 (2005).
\newblock \doi{10.17487/RFC4289}

\bibitem{rfc6838}
N.~Freed, D.J.C. Klensin, T.~Hansen.
\newblock {Media Type Specifications and Registration Procedures}.
\newblock RFC 6838 (2013).
\newblock \doi{10.17487/RFC6838}

\bibitem{Arai:2001}
H.~Arai, Sentence length and lognormal distribution : A case study of akutagawa
  and dazai.
\newblock Hitotsubashi Rev. \textbf{125}(3), 205--223 (2001).
\newblock {http://hermes-ir.lib.hit-u.ac.jp/rs/handle/10086/10418} (In
  Japanese)

\bibitem{Furuhashi:2012}
S.~Furuhashi, Y.~Hayakawa, Lognormality of the distribution of japanese
  sentence lengths.
\newblock J. Phys. Soc. Jpn. \textbf{81}(3), 034004 (2012).
\newblock \doi{10.1143/JPSJ.81.034004}

\bibitem{Sasaki:1976}
K.~Sasaki, Distribution of sentence-length.
\newblock Math. Linguist. \textbf{78}, 13--22 (1976).
\newblock (In Japanese)

\bibitem{Ishida:2007}
M.~Ishida, K.~Ishida, On distributions of sentence lengths in {Japanese}
  writing.
\newblock Glottometrics \textbf{15}, 28--44 (2007).
\newblock \urlprefix\url{https://api.semanticscholar.org/CorpusID:12215774}

\bibitem{borbely2019sentencelength}
G.~Borb\'{e}ly, A.~Kornai.
\newblock Sentence length (2019).
\newblock \doi{10.48550/arXiv.1905.09139}

\bibitem{Liu_Yang_Liu_2024}
J.~Liu, N.~Yang, H.~Liu, Distribution of sentence length of english complex
  sentences.
\newblock Moderna Spr\r{a}k \textbf{118}(3), 51–69 (2024).
\newblock \doi{10.58221/mosp.v118i3.15574}

\bibitem{ASI:ASI21426}
S.~Milojevi\'{c}, Power law distributions in information science: Making the
  case for logarithmic binning.
\newblock J. Am. Soc. Inf. Sci. Technol. \textbf{61}(12), 2417--2425 (2010).
\newblock \doi{10.1002/asi.21426}

\end{thebibliography}

\end{document}